\documentclass[article]{jss}
\usepackage[utf8]{inputenc}
\usepackage[english]{babel}
\usepackage{subcaption}
\usepackage{amsmath, array}
\usepackage{amsfonts}
\usepackage{bbm}
\usepackage{tikz}
\usetikzlibrary{automata,positioning}
\usepackage{bm}

\DeclareUnicodeCharacter{00A0}{ }

\author{Michael J. Kane\\Yale University \And
        Nan Chen\\The University \\ of Texas\And
        Alexander M. Kaizer\\University of Colorado\AND
        Xun Jiang\\Amgen Inc. \And
        H. Amy Xia\\Amgen Inc.\And
        Brian P. Hobbs\\Cleveland Clinic}
\title{Analyzing Basket Trials under Multisource Exchangeability Assumptions}

\Plainauthor{Michael J. Kane, First Author} %% comma-separated
\Plainauthor{Nan Chen, First Author} %% comma-separated
\Plainauthor{Alexander M. Kaizer, Second Author} %% comma-separated
\Plainauthor{Xun Jiang, Second Author} %% comma-separated
\Plainauthor{H. Amy Xia, Second Author} %% comma-separated
\Plainauthor{Brian P. Hobbs, Second Author} %% comma-separated
\Plaintitle{Basket Trial Analysis with Multisource Exchangeability Models} %% without formatting
\Shorttitle{Basket Trial Analysis with MEM Models} %% a short title (if necessary)

\Abstract{
Basket designs are prospective clinical trials that are devised with the hypothesis that the presence of selected molecular features determine a patient's subsequent response to a particular ``targeted'' treatment strategy. %Central to the design are assumptions 1) that a patient's expectation of treatment benefit can be ascertained from accurate characterization of their molecular profile and 2) that biomarker-guided treatment selection supersedes traditional clinical indicators for the studied populations, such as primary site of origin or histopathology. Thus, 
Basket trials are designed to enroll multiple clinical subpopulations to which it is assumed that the therapy in question offers beneficial efficacy in the presence of the targeted molecular profile. The treatment, however, may not offer acceptable efficacy to all subpopulations enrolled. Moreover, for rare disease settings, such as oncology wherein these trials have become popular, marginal measures of statistical evidence are difficult to interpret for sparsely enrolled subpopulations. Consequently, basket trials pose challenges to the traditional paradigm for trial design, which assumes inter-patient exchangeability. The R-package \pkg{basket} facilitates the analysis of basket trials by implementing multi-source exchangeability models. By evaluating all possible pairwise exchangeability relationships, this hierarchical modeling framework facilitates Bayesian posterior shrinkage among a collection of discrete and pre-specified subpopulations. %, allowing an arm to ``borrow'' power from other similar arms, depending on the extent to which the arms are exchangeable in the Bayesian sense. 
Analysis functions are provided to implement posterior inference of the response rates and all possible exchangeability relationships between subpopulations. 
In addition, the package can identify ``poolable'' subsets of and report their response characteristics. The functionality of the package is demonstrated using data from an oncology study with subpopulations defined by tumor histology.}
\Plainkeywords{Bayesian analysis, basket design, hierarchical model, master protocol, oncology, patient heterogeneity} %% without formatting
\Address{
  Michael J. Kane\\
  School of Public Health\\
  Biostatistics Department\\
  Yale University\\
  New Haven, CT, USA\\
  E-mail: \email{michael.kane@yale.edu}\\
  \\
  Nan Chen\\
  MD Anderson Cancer Center\\
  The University of Texas\\
  Houston, TX, USA\\
  E-mail: \email{nchen2@mdanderson.org}\\
  \\
  Alex Kaizer\\
  Colorado School of Public Health\\
  Department of Biostatistics and Informatics\\
  University of Colorado-Anschutz Medical Campus\\
  Aurora, CO, USA\\
  E-mail: \email{alex.kaizer@cuanschutz.edu}\\
  \\
  Xun Jiang\\
  Amgen Inc.
  Thousand Oaks, CA, USA\\
  E-mail: \email{xunj@amgen.com}\\
  \\
  H. Amy Xia\\
  Biostatistics and Design \& Innovation\\
  Amgen Inc.
  Thousand Oaks, CA, USA\\
  E-mail: \email{hxia@amgen.com}\\
  \\
  Brian Hobbs\\
  Taussig Cancer Center\\
  The Cleveland Clinic\\
  Cleveland, OH, USA\\
  E-mail: \email{HobbsB@ccf.org}
}
\begin{document}

%% include your article here, just as usual
%% Note that you should use the \pkg{}, \proglang{} and \code{} commands.

\section[intro]{Introduction}
%% Note: If there is markup in \(sub)section, then it has to be escape as above.

%\makeatletter{\renewcommand*{\@makefnmark}}{}
Basket designs are prospective clinical trials that are devised with the hypothesis that the presence of selected molecular features determine a patient's subsequent response to a particular ``targeted'' treatment strategy. Central to the design are assumptions 1) that a patient's expectation of treatment benefit can be ascertained from accurate characterization of their molecular profile and 2) that biomarker-guided treatment selection supersedes traditional clinical indicators for the studied populations, such as primary site of origin or histopathology. Thus, basket trials are designed to enroll multiple clinical subpopulations to which it is assumed that the therapy(s) in question offers beneficial efficacy in the presence of the targeted molecular profile(s). %A basket trial is an uncontrolled, multi-arm clinical trial measuring the effect of a single treatment over partitions of patients, usually based on their diseases or subtypes. 
These designs have become popular as drug developers seek to conform therapeutic interventions to the individuals being treated with precision medicine and biomarker-guided therapies.
%Since they group patients, acknowledging the  heterogeneity of treatment benefit among patients and patient subpopulations, they have become popular in precision medicine which seeks to conform therapeutic interventions to the individuals being treated. %Basket trial design is becoming increasingly popular, especially in oncology trials evaluating the effectiveness of a therapeutic strategy among patients defined by the presence of a particular drug target (often a genetic mutation) rather than a particular tumor histology.
Most basket trials have been conducted within exploratory settings to evaluate agent-specific estimates of tumor response. Cunanan et al.\ \citep{doi:10.1200/JCO.2016.69.9751} describe three studies implemented in oncology settings which extend the basic formulation of a basket trial to multiple targets and/or agent combinations. Most commonly uncontrolled trials, extensions have recently accommodated a wide variety of potential motivations beyond exploratory studies. 

Molecularly targeted treatment strategies may not offer acceptable efficacy to all putatively promising clinical indications. Early basket trials were criticized for their reliance on basketwise analysis strategies that suffered from limited power in the presence of imbalanced enrollment as well as failed to convey to the clinical community evidentiary measures of heterogeneity among the studied clinical subpopulations, or ``baskets''. Acknowledging the potential for differential effectiveness among the enrolled patient subpopulations by design, heterogeneity exists as an intrinsic hypothesis in  evaluations of treatment efficacy. Moreover, for rare disease settings, such as oncology wherein these trials have become popular, marginal measures of statistical evidence are difficult to interpret on the basis of individual basket-wise analyses for sparsely enrolled subpopulations. Consequently, basket trials pose specific challenges to the traditional paradigm for trial design, which assume that the patients enrolled represent a statistically exchangeable cohort.
%are conducted with respect to arms or ``basket'' which collectively represent a partition of the targeted patient population consisting of discrete subtypes. Acknowledging the potential for differential effectiveness on the basis of traditional criteria for cancer subtyping, evaluations of treatment effectiveness are conducted with respect to arms or ``basket'' which collectively represent a partition of the targeted patient population consisting of discrete subtypes.

%hobbs2018monitor
\cite{hobbs2018monitor} extended the Bayesian multisource exchangeability model (MEM) framework to basket trial design and subpopulations inference. Initially proposed by \cite{kaizer2017}, the MEM framework addressed the limitations associated with ``single-source'' Bayesian hierarchical models, which rely on a single parameter to determine the extent of influence, or shrinkage, from all sources. 
In the presence of subpopulations that arise as mixtures of exchangeable and non-exchangeable subpopulations, single-source hierarchical models (SEM) are characterized by limited borrowing, even in the absence of heterogeneity \citep{kaizer2017}. Moreover, when considering the effectiveness of a particular treatment strategy targeting a common disease pathway that is observed among differing histological subtypes, SEMs fail to admit statistical measures that delineate which patient subtypes should be considered ``non-exchangeable'' based on the observed data. 
%produces shrinkage estimators that can be used as the basis for integrating supplementary data into the analysis of a primary data source. This approach addressed the previous limitation associated with Bayesian hierarchical models requiring prespecification of a shrinkage weight for each source or relies on the data to inform a single parameter, which determines the extent of influence or shrinkage from all sources, risking considerable bias or minimal borrowing. 
By way of contrast, MEM provides a general Bayesian hierarchical modeling strategy accommodating source-specific smoothing parameters. MEMs yield multi-resolution smoothed estimators that are asymptotically consistent and accommodate both full and non-exchangeability among discrete subpopulations.
%that is asymptotically consistent while reducing the dimensionality of the prior space. approach for integrating multiple, potentially non-exchangeable, supplemental data sources into the analysis of a primary data source. The modeling framework yields source-specific smoothing parameters that can be estimated in the presence of the data to facilitate a dynamic multi-resolution smoothed estimator that is asymptotically consistent while reducing the dimensionality of the prior space. 
The inclusion of methods for shrinkage of multiple sources is not restricted to use in basket trial master protocols, but has also been extended in the MEM framework to a sequential combinatorial platform trial design where it demonstrated improved efficiency relative to approaches without information sharing \citep{kaizer2018}.

This paper introduces the \pkg{basket} \citep{basket} package for the R-programming environment \citep{Rcore} to analyze basket trials under MEM assumptions. The main analyses conduct full posterior inference with respect to a set of response rates corresponding to the studied subpopulations. The posterior exchangeability probability (PEP) matrix is calculated, which describes the probability that any pair of baskets are exchangeable. Based on the resultant PEP, subpopulations are clustered into meta-baskets.  Additionally, posterior effective sample sizes are calculated for each basket, describing the extent of posterior shrinkage achieved. Posterior summaries are reported for both ``basketwise'' and ``clusterwise'' analyses.

The package used in the examples below is available on CRAN at \url{https://cran.r-project.org/package=basket} and it fits into the general category of the ``Design and Analysis of Clinical Trials'' \citep{clinicaltrialstask} focusing on uncontrolled, early-phase trial analysis. The interface is designed to be simple and will readily fit into clinical trial frameworks. It has been tested using \proglang{R} version 3.5 and the \pkg{basket} package version 1.0.0.%, including \pkg{crmPack} \citep{crmPack}.

%of two-stage (or doubly) randomized clinical trials for both the original \cite{Rucker:1989aa} version (unstratified) and the stratified design by \cite{Cameron:2016aa}. This package facilitates the implementation of the two-stage randomized design by providing the necessary sample size estimation and analysis tools for clinicians and statisticians seeking to disentangle the roles of patient preference in clinical trials.  The package is available on CRAN \citep{cameron2018}.

\section{Exchangeability for Trials with Subpopulations}

\subsection{The Single-Source Exchangeability Model}

\begin{figure}[htbp!]
\centering
\begin{tikzpicture}[shorten >=1pt,node distance=2cm,on grid,auto]
  \node[state] (y1)  {$Y_1$}; 
  \node[state] (y2) [right=of y1] {$Y_2$}; 
  \node[state] (y3) [right=of y2] {$Y_3$}; 
  \node (dots) [right=of y3] {$...$}; 
  \node[state] (yk) [right=of dots] {$Y_J$}; 
  \node[state] (theta1) [above=of y1]  {${\bm\theta_1}$}; 
  \node[state] (theta2) [right=of theta1] {${\bm\theta_2}$}; 
  \node[state] (theta3) [right=of theta2] {${\bm\theta_3}$}; 
  \node (dots2) [right=of theta3] {$...$}; 
  \node[state] (thetak) [right=of dots2] {${\bm\theta_J}$}; 
  \node[state] (theta) [above=of theta3] {${\bm\theta}$};
  \path[->]
  (y1) edge node {} (theta1)
  (y2) edge node {} (theta2)
  (y3) edge node {} (theta3)
  (yk) edge node {} (thetak)
  (theta1) edge node {} (theta)
  (theta2) edge node {} (theta)
  (theta3) edge node {} (theta)
  (thetak) edge node {} (theta);
\end{tikzpicture}
\caption{A conventional single-source Bayesian hierarchical model with $J$ subtypes.} 
\label{slbhm}
\end{figure}
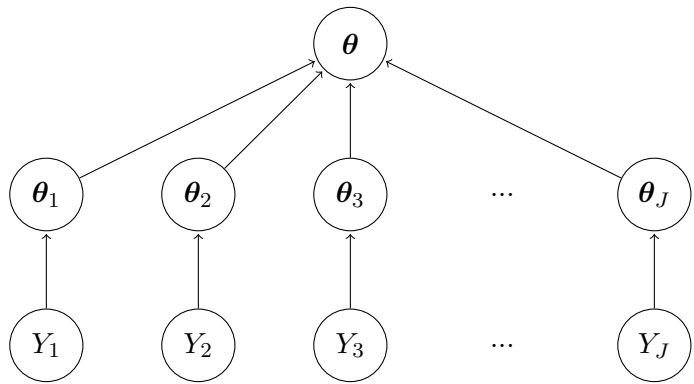

%   \node[state,initial] (q_0)   {$q_0$}; 
%   \node[state] (q_1) [above right=of q_0] {$q_1$}; 
%   \node[state] (q_2) [below right=of q_0] {$q_2$}; 
%   \node[state,accepting](q_3) [below right=of q_1] {$q_3$};
%    \path[->] 
%    (q_0) edge  node {0} (q_1)
%          edge  node [swap] {1} (q_2)
%    (q_1) edge  node  {1} (q_3)
%          edge [loop above] node {0} ()
%    (q_2) edge  node [swap] {0} (q_3) 
%          edge [loop below] node {1} ();

Basket trials intrinsically include subpopulations, which require {\it a priori} consideration for inference. When ignored the trial simply pools patients, conducting inference with the implicit assumption of inter-patient statistical exchangeability, which can induce bias and preclude the identification of unfavorable/favorable subtypes in the presence of heterogeneity. At the other extreme, subpopulation-specific analyses assume independence. While attenuating bias, this approach suffers from low power, especially in rare subpopulations enrolling limited sample size. Bayesian hierarchical models address this polarity, facilitating information sharing by ``borrowing strength'' across subtypes with the intent of boosting the effective sample size of inference for individual subtypes.

\textit{Single-source exchangeability models} (SEM), represent one class of Bayesian hierarchical models. In the context of a basket trial design, statistical approaches using the SEM framework rely on a single parametric distributional family to characterize heterogeneity across all subpopulations, which is computationally tractable but intrinsically reductive in characterization of heterogeneity. In the presence of both exchangeable and non-exchangeable arms, the SEM framework tends to favor the extremes of no borrowing or borrowing equally from all sources, effectively ignoring disjointed singleton subpopulations and meta-subtypes. % For the oncology clinical trials setting, the data may be responses to a treatment grouped histology and the latent parameters to be fit may give the distribution of the overall treatment effect. The model could shrink the group-wise estimates toward the overall mean based on the cohort size along with other model parameters. The posterior model estimates will be more accurate and incur less risk because they are able to use information, or ``borrow power'' from other cohorts.

%but intrinsically reductive in their characterization of heterogeneity.
Consider a basket trial which enrolls patients from $J$ subpopulations (or subtypes) ($j=1,...,J$), where $Y_j$ represents the responses observed among patients in the $j$th subtypes. Using $i$ to index each patient, the SEM generally relies on model specifications that assume that patient-level responses, $Y_{i,j}$, are exchangeable Bernoulli random variables conditional on subtype-specific model parameters, e.g.\ ${\bm\theta_j}$. The second-level of the model hierarchy assumes that the collection of subtype-specific model parameters, ${\bm\theta_1},$ $...$ $,{\bm\theta_J},$ are statistically exchangeable through the specification of a common parent distribution. Figure \ref{slbhm} illustrates this structure, wherein each $Y_j$ has its own subtype-specific ${\bm\theta_j}$ which are further assumed exchangeable to estimate the overall ${\bm\theta}$. 
Examples of SEM approaches are introduced and discussed by \cite[][chapter 2]{berry:2010}, \cite{thall2003sim}, and \cite{berry2013ct}, with \cite{hobbs2018monitor} providing additional background on these specific SEM implementations. SEM approaches are also implemented in packages by \cite{bclust} and \cite{BHC} and have been extended to more specialized applications in fMRI studies \citep{stocco2014}, modeling clearance rates of parasites in biological organisms \citep{bhrcr}, modeling genomic bifurcations \citep{mfa}, modeling ChIP-seq data through hidden Ising models \citep{iSeq}, modeling genome-wide nucleosome positioning with high-throughput short-read data \citep{RJMCMCNucleosomes}, and modeling cross-study analysis of differential gene expression \citep{XDE}.

While integrating inter-cohort information, SEMs are limited by assumptions of exchangeability among all cohorts. That is, the joint distribution $\mathbb{P}(Y_1, Y_2, ..., Y_k)$ is invariant under a permutation describing subpopulation subsets. $\mathbb{P}(Y_1, Y_2, ..., Y_k) = \mathbb(Y_k, ..., Y_2, Y_1)$. SEMs are ``single-source'' in the sense that the model uses a single set of parameters to characterize heterogeneity such that the statistical exchangeability of model parameters is always assumed. Violations of these assumptions with analyses of response rates in clinical trials yields bias, potentially inflating the estimated evidence of an effective response rate for poorly responding cohort or minimizing the effect in effective subsets. These assumptions have resulted in poor results for frequentist power when controlling for strong type I error, leading some cancer trialists to question the utility of Bayesian hierarchical models for phase II trials enrolling discrete subtypes \citep{freidlin2013ccr,Cunananetal17specifying}.

\subsection{The Multi-source Exchangeability Model}

Limitations of SEM can be overcome through model specification devised to explicitly characterize the evidence for exchangeability among collections of subpopulations enrolling in a clinical trial. %relaxing the constraint on a single-source and refining the hierarchical model specification to flexibly consider multi-sources which relax the assumption of inter-cohort data exchangeability and generally require prespecification of the amount of borrowing under different paradigms related to the power prior. 
Multi-source exchangeability models (MEM) produce cohort-specific smoothing parameters that can be estimated in the presence of the data to facilitate dynamic multi-resolution smoothed estimators that reflect the extent to which subsets of subpopulations should be consider exchangeable. Shown to be asymptotically consistent, MEMs were initially proposed by \cite{kaizer2017} for ``asymmetric'' cases wherein a primary data source is designated for inference in the presence of potentially non-exchangeable supplemental data sources. %is integrated into the inference of a primary source. 
The framework was extended by \cite{hobbs2018monitor} to the ``symmetric'' case wherein no single source or subtype is designated as primary (e.g., a basket trial). The symmetric MEM approach considers all possible pairwise exchangeability relationships among $J$ subpopulations and estimates the probability that any subset of subpopulations should be considered statistically exchangeable (or poolable). %\citep[see e.g.][]{kaizer2017}.

The symmetric MEM is the motivation and focus of the \pkg{basket} package. 
%The MEM provides a hierarchical specification to explicitly identify exchangeability among baskets. 
While SEMs are parameterized by a single set of parameters ${\bm \theta}$, the MEM may have up to $J$ (the number of subtypes) sources of exchangeability with each set of data $Y_j$ contributing to only one set of parameters.  All possible combinations of exchangeability can be enumerated, denoted as $K$ possible configurations ($\Omega_k$, $k=1,...,K$).

\begin{figure}[htbp!]
\centering
  \begin{subfigure}[b]{0.45\textwidth}
    \centering
\begin{tikzpicture}[shorten >=1pt,node distance=2cm,on grid,auto]
  \node[state] (y1)  {$Y_1$}; 
  \node[state] (y2) [right=of y1] {$Y_2$}; 
  \node[state] (y3) [right=of y2] {$Y_3$}; 
  \node[state] (theta_1) [above=of y2] {${\bm\theta_1}$};
  \node[state] (theta_2) [above=of y3] {${\bm\theta_2}$};
  \path[->]
  (y1) edge node {} (theta_1)
  (y2) edge node {} (theta_1)
  (y3) edge node {} (theta_2);
\end{tikzpicture}    \caption{Model where $Y_1$ and $Y_2$ are exchangeable.}
    \label{cluster_post_density_graph}
  \end{subfigure}
  \begin{subfigure}[b]{0.45\textwidth}
    \centering
\begin{tikzpicture}[shorten >=1pt,node distance=2cm,on grid,auto]
  \node[state] (y1)  {$Y_1$}; 
  \node[state] (y2) [right=of y1] {$Y_2$}; 
  \node[state] (y3) [right=of y2] {$Y_3$}; 
  \node[state] (theta_1) [above=of y2] {${\bm\theta_1}$};
  \node[state] (theta_2) [above=of y3] {${\bm\theta_2}$};
  \path[->]
  (y1) edge node {} (theta_1)
  (y2) edge node {} (theta_2)
  (y3) edge node {} (theta_1);
\end{tikzpicture}    \caption{Model where $Y_1$ and $Y_3$ are exchangeable.}
    \label{post_density_graph}
  \end{subfigure}
  \caption{Two example exchangeability configurations of the MEM.}
  \label{mem_config}
\end{figure}
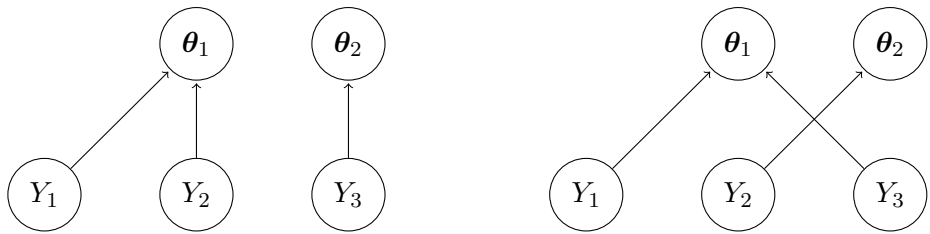

\subsubsection{Model Description}

Figure \ref{mem_config} depicts two possible MEMs among three subpopulations wherein at least two subpopulations are statistically exchangeable. Both examples comprise two ``sources'' of exchangeability for inference, with $Y_1$ and $Y_2$ combined to represent one ``source'' to estimate ${\bm\theta_1}$ and $Y_3$ to estimate ${\bm\theta_2}$ in (a) and $Y_1$ and $Y_3$ combined in (b). Implementation of \pkg{basket} considers the number of ``sources'' ranging from one (as in the single-source case), wherein all subtypes are pooled together, to $J$, the total number of subtypes. The MEM Bayesian model specification facilitates posterior inference with respect to all possible pairwise exchangeability relationships among $J$ subpopulations. The framework facilitates estimation of disjointed subpopulations comprised of meta-subtypes or singelton subtypes and thereby offers additional flexibility when compared to SEM specifications.

%To construct the model, we start by considering the total set of MEMs. The model will be restricted so that a data set $Y_j$ will be used to fit only one set of parameters. However, later it will be shown that the analysis provides a measure of exchangeability between any two baskets. The number of ``sources'' can range from one (as in the single-source case), wherein all subtypes are pooled together, to $J$, the total number of subtypes. In Figure \ref{mem_config} both multi-source exchangeability models have 3 subtypes but represent two different examples where there are 2 ``sources'' being used for inference, with $Y_1$ and $Y_2$ combined to represent one ``source'' to estimate ${\bm\theta_1}$ and $Y_3$ to estimate ${\bm\theta_2}$ in (a) and $Y_1$ and $Y_3$ combined in (b).
%
The set space of all possible pairwise exchangeability relationships among a collection of $J$ discrete cohorts can be represented by a symmetric $J \times J$ matrix $\bm{\Omega}$ with element $\Omega_{ij} = \Omega_{ji} \in [0, 1]$ with value 1 (0) indicating that patients of subtype $i$ are statistically exchangeable with (independent of) patients of subtype $j$. Without additional patient-level characteristics, it is assumed patients within an identical subtype are assumed to be statistically exchangeable. That is $\Omega_{ii} = 1$ for $\{i : 1, ..., J\}$. There are $K = \prod_{j=1}^{J-1} 2^j$ possible configurations of $\bm{\Omega}$, each representing one possible pairwise exchangeability relationship among the $J$ subtypes. The framework differs fundamentally from SEM in that it allows for the existence of multiple closed subpopulations (or cliques) comprised of fully exchangeable subtypes. Therefore, following the terminology of \cite{kaizer2017} we refer to each possible configuration of $\bm{\Omega}$ as a MEM. 

For a basket trial designed to enroll a total of $N$ patients in $J$ baskets, let $y_{ij} = 1$ indicate the occurrence of a successful response for the $i$th patient enrolled in basket $j$, and 0 indicate treatment failure. Let $n_j$ denote the number of patients observed in basket $j$ and denote the total number of responses in basket $j$ by $S_j = \sum_{i=1}^{n_j} y_{ij}$. The set $\{S_1, S_2, ..., S_J\}$ is denoted ${\bm S}$.  Let ${\bm \pi} = \{\pi_1, \pi_2, ... \pi_J\}$ vectorize the set of response rates such that $\pi_j$ denotes the probability of response for $j$th basket and $S_j \sim $Bin($n_j, \pi_j$) with prior distribution $\pi_j \sim$ Beta($a_j$, $b_j$). Let $B()$ denote the beta function. Given an exchangeability configuration $\bm{\Omega}_j,$ the marginal density of $S_j$ follows as \citep[see][for details]{hobbs2018monitor}
%obtained by integrating the likelihood of $\pi_j\: |\: S_j$ with respect to $p( \bm{\pi}\: |\:  \bm{S}_{(-j)}),$ 
\begin{equation} \begin{split}\label{margData}
m(\bm{S}_j\:|\: \bm{\Omega}_{j},\: \bm{S}_{(-j)}) \propto \frac{B\left(
a + \sum_{h=1}^{J} \Omega_{j,h} S_{h} ,\: b + \sum_{k=1}^{J} \Omega_{j,k}(n_{k} - S_{k}
\right)}{B(a,b)} \times \\ 
\prod_{i=1}^{J}  \left(\frac{B(a + S_{i},\: b + n_{i} - S_{i})}{B(a,b)}\right)^{1-\Omega_{j,i}}.
\end{split} \end{equation}

Marginal posterior inference with respect to $\pi_j$ $|$ $\bm{S}$ averages the conditional posterior of $\pi_j$ $|$ $\bm{\Omega}_{j},$ $\bm{S}$  with respect to the marginal posterior probability of $G=2^{J-1}$ possible exchangeability configurations of $\bm{\Omega}_j.$ Let $\bm{\omega}$ $=$ $\{\bm{\omega}_1,$ $...,$ $\bm{\omega}_G\}$ denote the collection of vectors each of length $J$ that collectively span the sample space of $\bm{\Omega}_j.$ The marginal posterior distribution can be represented by a finite mixture density
\begin{equation} \label{margPost}
q(\pi_{j}|\bm{S}) \propto \sum_{g=1}^{G} q(\pi_j\: |\: \bm{S}, \bm{\Omega}_j=\bm{\omega}_g)Pr(\bm{\Omega}_j=\bm{\omega}_g\: |\: \bm{S}),
\end{equation}
where the posterior probability of exchangeability configuration $\bm{\omega}_g$ given the observed data follows from Bayes' Theorem in proportion to the marginal density of the data given $\bm{\omega}_g$ and its unconditional prior probability %of $\bm{\omega}_g$
\begin{equation} \label{modelProb}
Pr(\bm{\Omega}_j=\bm{\omega}_g\: |\: \bm{S}) \propto \frac{m(\bm{S}_j\:|\: \bm{\Omega}_{j}=\bm{\omega}_g,\: \bm{S}_{(-j)}) Pr(\bm{\Omega}_{j}=\bm{\omega}_g)}{\sum_{u=1}^{G} m(\bm{S}_j\:|\: \bm{\Omega}_{j}=\bm{\omega}_u,\: \bm{S}_{(-j)}) Pr(\bm{\Omega}_{j}=\bm{\omega}_u) }.
\end{equation}
Model specification for the symmetric MEM method is described in detail by \cite{hobbs2018monitor}.

\subsubsection{Estimating Basketwise Exchangeability}

The \pkg{basket} package computes the posterior probability that subpopulations $i$ and $j$ should be considered statistically exchangeable. The collection of all pairwise posterior exchangeability probabilities (PEP) is denoted in the output as the PEP matrix. Additionally, \pkg{basket} identifies the maximum {\em a posteriori} (MAP) multisource exchangeability model.

Let $\mathcal{O}$ denote the entire sample domain of $\bm{\Omega}$ comprised of $K$ $=$ $\prod_{j=1}^{J-1} 2^j$ strictly symmetric MEMs. The PEP matrix is obtained by evaluating the union of MEMs for which $\Omega_{ij} = 1$ over the sample domain of $\mathcal{O}$,
\begin{equation*}
    \mathbb{P}(\Omega_{ij} = 1 | {\bm S} ) = \sum_{\bm{\Omega} \in \mathcal{O}} \mathbbm{1}_{\{\Omega_{ij} = 1\}} \ \mathbb{P}(\bm{\Omega} | {\bm S}),
\end{equation*}
where $\mathbb{P}(\bm{\Omega} | {\bm S})$ is the product of row-wise calculations specified in Equation \ref{modelProb}. Note that there are $K$/2 MEM configurations in the space of $\mathcal{O}$ where $\Omega_{ij}=1$. The MAP follows as the MEM configuration that attains maximum $Pr(\bm{\Omega}\:|\: \bm{S})$ over $\mathcal{O}.$

\subsubsection{Effective Sample Size}

Measurement of the extent to which information has been shared across sources in the context of a Bayesian analysis is best characterized by the effective sample size (ESS) of the resultant posterior distribution \cite{hobbs2013, murray2015}. ESS quantifies the extent of information sharing, or Bayesian ``shrinkage,'' as the number of samples that would be required to obtain the extent of posterior precision achieved by the candidate posterior distribution when analyzed using a vague ``reference'' or maximum entropy prior. Calculation of the ESS in \pkg{basket} deviates from the approach suggested in \cite{hobbs2018monitor}, which is sensitive to heavy-tailed posteriors. Robustness is introduced with \pkg{basket} through beta distributional approximation, which yields more conservative estimates of ESS. Specifically, the simulated annealing algorithm (implemented with \code{GenSA} package \cite{gensa}) is used to identify the parametric beta distribution with minimal Euclidean distance between the interval boundaries obtained from the posterior estimated HPD interval and the corresponding beta $1-$\code{hpd_alpha} Bayesian credible interval. Shape parameters attained from the ``nearest'' parametric beta distribution are summed to yield estimates of posterior ESS for each basket and cluster.
%\textbf{In the package, two methods (the exact method and the MCMC method) are developed and more details regarding two methods can be found from the next section. The ESS calculation in two methods are different. In the exact method, the entire MEM sample space are enumerated and the ESS values are calculated from the weighted average of the $P(\Omega)$ and the corresponding $ESS(\Omega)$, which can be obtained from the above derivation. In the MCMC method, the ESS is calculated through the highest probability density (HPD) interval matching method. The posterior distribution of response rate of each basket is sampled based on the MCMC method and the HPD interval is subsequently calculated from the sampled posterior distribution. The HPD interval of the effective posterior distribution theoretically should be the same as that from the sampled posterior distribution. The ESS is calculated by minimizing the difference between the HPD intervals of the effective distribution and the sampled posterior distribution. We used the simulated annealing method for the minimizing computation. }      
%Given a specific MEM, the ESS of the conditional posterior is
%\begin{equation}
%    \text{ESS}(\bm{\Omega}_j) = a + b + \sum_{i = 1}^J \Omega_{ji} n_i .
%\end{equation}

\subsubsection{Posterior Probability}

Basket trials are devised for the purpose of testing the hypothesis that a targeted treatment strategy achieves sufficiently promising activity among a partition of the targeted patient population. The MEM framework acknowledges the potential for heterogeneity with respect to the effectiveness of the enrolled patient subpopulations or baskets. Within the MEM framework, this testing procedure follows from the cumulative density function (cdf) of the marginal posterior distribution (\ref{margPost}). Specifically, the posterior probability that $\pi_j$ exceeds a null value $\pi_0$ is computed by the weighted average of cdfs for all possible exchangeability configurations. \pkg{Basket} implements this computation and allows for subpopulation-specific values of the null hypothesis, $\pi_0$, which quantify differing benchmarks for effectiveness among the studied baskets. Note that this feature accommodates basket formulation on the basis of varying levels of clinical prognosis.
%Evaluating the hypothesis that the response probability for a targeted intervention exceeds a null value, denoted $\pi_0$, while acknowledging the potential for heterogeneity in effectiveness in accordance with the prespecified basket partitions is accomplished within the MEM framework by finding the probability that $\pi_j$ exceeds $\pi_0$. This value is calculated as the weighted average of cumulative distributions functions for all possible exchangeability configurations,
%\begin{equation*} \begin{split} \mathbb{P}(\pi_{j}\:>\:\pi_0\: |\: \bm{S}) = \sum_{g=1}^{G} \mathbb{P}(\bm{\Omega}_j=\bm{\omega}_g\: |\: \bm{S})\left[ 1-\frac{\int_{0}^{\pi_0}u^{a + \sum_{h=1}^{J} \omega_{g,h} S_{h}-1}(1-u)^{b + \sum_{k=1}^{J} \omega_{g,k}(n_{k} - S_{k})-1}du}{B(a + \sum_{h=1}^{J} \omega_{g,h} S_{h},\: b + \sum_{k=1}^{J} \omega_{g,k}(n_{k} - S_{k})-1)}\right]. \end{split} \end{equation*}

\section{Package Overview}

The \pkg{basket} package facilitates implementation of the binary, symmetric multi-source exchangeability model with posterior inference arising with both exact computation and Markov chain Monte Carlo sampling. The user is required to input vectors that describe the number of samples (\code{size}) and observed successes (\code{responses}) corresponding to each subpopulation (or basket). Analysis output includes full posterior samples, highest posterior density (HPD) interval boundaries, effective sample sizes (ESS), mean and median posterior estimates, posterior exchangeability probability matrices, and the maximum \emph{a posteriori} MEM. In addition to providing ``basketwise'' analyses, the package includes similar calculations for ``clusterwise'' analyses for which subgroups are combined into meta-baskets, or clusters, using a graphical clustering algorithm implemented with the \code{igraph} package that treats the posterior exchangeability probabilities as edge weights. A specific clustering algorithm is specified via argument \code{cluster_function} which is set to \code{"cluster_louvain"} by default. In addition, plotting tools are provided to visualize basket and cluster densities as well as their exchangeability.

Analysis requires the specification of beta shape parameters (\code{shape1} and \code{shape2}) for the prior distributions of the basketwise response probabilities $\pi_j.$  Shape parameter arguments may be specified as single positive real values, by which identical prior distributions are assumed for all $\pi_j,$ or as vectors of length $J$ with each pair of \code{shape1} and \code{shape2} values corresponding to each basket. Arguments \code{shape1} and \code{shape2} assume values $0.5$ by default characterizing prior distributions with the effective sample size of 1 patient for each $\pi_j.$

The user must additionally specify the symmetric matrix of prior exchangeability probabilities (\code{prior}). The model assumes that exchangeable information is contributed among patients enrolling into a common basket. Thus, all diagonal entries of \code{prior} must assume value 1. Off-diagonal entries, however, quantify the \emph{a priori} belief that each pair of subpopulations represents an exchangeable unit. Thus, off-diagonal cells of \code{prior} may assume any values on the unit interval. \pkg{Basket} assumes the ``reference'' prior proposed by \cite{hobbs2018monitor} as the default setting for which all off-diagonal cells assume prior probability 0.5, and thus are unbiased with respect to exchangeability in the absence of the data.

Evidence for sufficient activity is reported by basket and cluster as posterior probabilities. Posterior probability calculations require the further specification of either a null response rate or vector of null response rates corresponding to each basket (\code{p0} set to $0.15$ by default) as well as the direction of evaluation (\code{alternative} set to ``greater'' by default). Additionally, summary functions report the posterior estimates by basket and cluster. The highest posterior density (HPD) is calculated for a given a level of probabilistic significance (\code{hpd_alpha} set to 0.05 by default).

Bayesian computation is implemented by two methods: the exact method (\code{mem_exact()} function) and the Markov chain Monte Carlo (MCMC) sampling method (\code{mem_mcmc()} function). \code{mem_mcmc()} is the preferred method. \code{mem_exact()} provides slightly more precise estimates than the former but scales poorly in number of baskets. The discrepancy in precision between exact and sampling-based implementations is easily controlled by specifying a larger number of MCMC iterations (\code{num_iter} set to \code{2e+05} by default) in \code{mem_mcmc()}.

\subsection{The Exact Method and the MCMC Method}
Implementation of \code{mem_exact()} conducts posterior inference %of $\mathbb{P}(\bm{\Omega}_j = \bm{\omega}_g |\, {\bm S} )$ 
through enumeration of the entire sample domain of MEMs, denoted $\mathcal{O}$ above. Facilitating precise calculation of the posterior estimators, \code{mem_exact()} is computationally feasible only in the presence of a small number of subpopulations. %provides the exact accurate solution for trials with low number of baskets,  that yields a good estimation for trials with large number of baskets. In the exact method, the values of $\mathbb{P}(\bm{\Omega}_j = \bm{\omega}_g |\, {\bm S} )$ are calculated based on all $\bm{\Omega}$ configurations in the entire sample domain $\mathcal{O}$ in Equation \ref{eq_likelihood}, thus the solution is exactly accurate because the entire sample domain $\mathcal{O}$ are taken into account. 
Increasing the size of $J$ increases the number of configurations in $\mathcal{O}$ by order of $\mathcal{O}(2^{J^2}).$ Thus, the exact computation is impractical for large values of $J$. We recommend its use for $J < 7$.

Our MCMC sampling method, formulated from the Metropolis algorithm \citep[see e.g.][]{gelman:2013}, extends the model's implementation to larger collections of subpopulations, which currently accommodates more than $J=20$ baskets. Specifically, MCMC sampling is used to approximate the posterior distribution $\mathbb{P}(\bm{\Omega}_j = \bm{\omega}_g |\, {\bm S} )$. Implementation of \code{mem_mcmc()} requires the specification of an initial MEM matrix (\code{initial_mem}) used as the starting point for $\bm{\Omega}$ from which to initiate the Metropolis algorithm. Argument \code{initial_mem} is set to \code{round(prior - 0.001)} by default, which for the default setting of \code{prior} yields the identity matrix. 

The MCMC algorithm proceeds in iterative fashion with each step selecting a random number of cells of $\bm{\Omega}$ to flip from 0 to 1 or from 1 to 0 to produce a new candidate MEM which we denote $\bm{\Omega}^*$. Acceptance criteria for the candidate $\bm{\Omega}^*$ compares the marginal posterior density of $\bm{\Omega}^*$ and its unconditional prior distribution with respect to the last accepted MEM matrix configuration. Denote the sum of log marginal posterior density and prior distribution with new candidate MEM configuration by $D^*$ and previously accepted configuration by $D_0,$ respectively. If $D^*-D_0 \geq 0,$ the candidate configuration is accepted. Otherwise, the new configuration is accepted randomly with probability $\exp{(D^*-D_0)}$. For each sampled $\bm{\Omega}$ configuration, $\pi_j,$ is sample from its conditional posterior distribution for all $j=1,...,J.$ 

The algorithm initiates with a burn-in period (\code{mcmc_burnin} set to 50,000 by default). Discarding the burn-in samples, PEP calculation with \code{mem_mcmc()} evaluates the distribution of sampled MEMs, reporting for all basket pair combinations the proportion of samples that identify basket $i$ as exchangeable with basket $j.$ The MAP calculation reports the posterior mode or most frequently sampled MEM. Bayesian computation facilitated by \code{mem_mcmc()} scales MEM analyses to more than 20 baskets. Specification of the size of the MCMC iterations (\code{num_iter}) is pivotal to attaining precise estimates of the resultant posterior quantities. Our investigations support the default value of \code{2e+05} as a practical lower bound. In practice, one may gradually increase the number of the MCMC iterations until the resultant PEP matrix  converges to stable values.
%Accurate estimates of posterior quantities are attained when implemented with a sufficiently large number of sample iterations. %\subsubsection{MCMC Sampling} Thus, the size of the MCMC iterations is crucial to the precision of the results of the MCMC method. Because the MCMC sampling is conducted on the $\Omega$ space and the $\Omega$ matrix includes multiple elements (for instance, 15 elements for the $\Omega$ matrix corresponding to 6 subgroups), large number of MCMC iterations is required to obtain the converged results. The discrete value (0 or 1) of each element in the $\Omega$ makes the convergence diagnostics difficult. In practice, we can gradually increase the number of the MCMC iterations until the stable convergence of the PEP values is observed. 

\subsection{MEM Data Structure and Associated Methods}

\begin{table}
\centering
\begin{tabular}{|l|l|} \hline
{\bf Method} & {\bf Return Description}  \\ \hline 
\code{basket_pep} & Basketwise PEP matrix \\ \hline
\code{basket_map} & Basketwise maximum \emph{a posteriori} probability (MAP) matrix \\ \hline
\code{cluster_baskets} & Basket assignments for each cluster \\ \hline
\code{cluster_pep} & Clusterwise PEP matrix \\ \hline
\code{cluster_map} & Clusterwise MAP matrix \\ \hline
\end{tabular}
\caption{MEM model accessor functions.}
\label{tab_access}
\end{table}

Analysis functions \code{mem_mcmc()} and \code{mem_exact()} are parameterized almost identically, with the former requiring extra arguments that control the MCMC algorithm: the current seed (for reproducibility), the length of burn-in and number of MCMC iterations for computation of posterior quantities, and an initial MEM matrix from which to start the algorithm. Function arguments are specified with reasonable default values for implementation of either analysis type. Both functions return a common list data structure. Both are derived from an abstract S3 \code{"exchangeability_model"} class with concrete type \code{"mem_mcmc"} or \code{"mem_exact"} depending on which function generated the analysis. The two data structures differ only by extra elements included with \code{"mem_mcmc"} objects to control implementation of the MCMC algorithm. For convenience, and to promote using \code{"mem_mcmc"} by default, a wrapper function \code{basket()} was created. The \code{method} argument allows the user to specify the analysis function as either MCMC (via \code{"mcmc"}) or exact (via \code{"exact"}). By default the argument is set to \code{"mcmc"}.

MEM or ``exchangeability'' objects are composed of named elements. The first, \code{"call"} is the expression used to generate the analysis. Second is the \code{"basket"} element, which is a list with concrete class \code{mem_basket}, derived from the \code{mem} abstract class. Basket reports posterior estimates of trial subpopulations including the PEP, HPD interval, posterior probability, ESS, and other distribution characteristics. The \code{"cluster"} element comprises a list with concrete class \code{mem_cluster} and abstract class \code{mem} which contains posterior estimates for clusters rather than baskets. In addition to these three elements, an \code{mem_mcmc} object will also contain the seed used to generate the results. This value can be used to reproduce subsequent analyses.

Because they are relatively complex, a \code{summary} function is implemented to summarize the components relevant to exchangeability models for trial analysis. The \\
\code{summary.exchangeability_model()} method returns an object of type \code{"mem_summary"}. A \code{print.mem_summary()} method is provided for a user-readable summary of the trial. Because there is little distinction between an \code{exchangeability_model} object and its summary, \code{print.exchageability_model()} method prints the summary object.

The \code{mem_summary} object provides access to the overall study characteristic. Accessor methods are also provided to extract other key information from the analysis objects at both the basket and cluster levels. These functions and their descriptions are given in Table \ref{tab_access}.
In addition, a complete MEM analysis is computationally intensive; altering the null response rate need not imply rerunning the entire analysis. To facilitate  partial analysis updates under a new null (argument \code{p0}), the \code{update_p0()} function is provided. Likewise samples can be drawn from the posterior distribution of the basket and cluster models using the \code{sample_posterior()} function.

\subsection{Basketwise and Clusterwise Visualization}

Two types of functions are provided for visualizing the results of an MEM analysis, both of which are supported at basket and cluster levels of inference. Density plotting is available with the \code{plot_density()} functions, which produce graphs depicting the posterior distributions of response probabilities at the basket and cluster level. Additionally, functions for visualizing exchangeability relationships are provided in a manner similar to correlograms. Since the values visualized are exchangeability, rather than correlation, we have termed these plots {\em exchangeograms}. These can be plotted for PEP and MAP matrices using the \code{plot_pep()} and \code{plot_map()} functions, respectively.

\section{Case Study: The Vemurafenib Basket Trial}

The ``Vemurafenib in multiple nonmelanoma cancers with BRAF V600 mutations'' study \citep{hyman2015}, enrolled patients into predetermined baskets that were determined by organ site with primary end point defined by Response Evaluation Criteria in Solid Tumors (RECIST), version 1.1 \citep{eisenhauer2009} or the criteria of the International Myeloma Working Group (IMWG) \citep{durie2006}. Statistical evidence for preliminary clinical efficacy was obtained through estimation of the organ-specific objective response rates at 8 weeks following the initiation of treatment. This section demonstrates the implementation of \pkg{basket} through analysis of six organs comprising non–small-cell lung cancer (NSCLC), cholangiocarcinoma (Bile Duct), Erdheim–Chester disease or Langerhans’-cell histiocytosis (ECD/LCH), anaplastic thyroid cancer (ATC), and colorectal cancer (CRC) which formed two cohorts. Patients with CRC were initially administered vemurafenib. The study was later amended to evaluate vemurafenib in combination with cetuximab for CRC which comprised a new basket. Observed outcomes are summarized in Table \ref{tab_vemu} by basket. Included in the \pkg{basket} package, the dataset is accessible in short \code{vemu_wide} as well as long formats \code{vemu}.

\begin{table}
\centering
\begin{tabular}{|l|l|l|l|l|} \hline
{\bf Basket} & {\bf Enrolled} & {\bf Evaluable} & {\bf Responses} & {\bf Response Rate}\\ \hline 
NSCLC          & 20 & 19 & 8  & 0.421 \\ \hline 
CRC (vemu)     & 10 & 10 & 0  & 0.000   \\ \hline 
CRC (vemu+cetu) & 27 & 26 & 1  & 0.038 \\ \hline 
Bile Duct      & 8  &  8 & 1  & 0.125 \\ \hline 
ECD or LCH     & 18 & 14 & 6  & 0.429 \\ \hline 
ATC            & 7  &  7 & 2  & 0.286 \\ \hline 
\end{tabular}
\caption{Vemurafenib trial enrollment and responses.}
\label{tab_vemu}
\end{table}

Inspection of Table \ref{tab_vemu} reveals heterogeneity among the studied baskets. CRC (vemu), CRC (vemu+cetu), and Bile Duct had relatively low response rates when compared to other baskets, suggesting that patients presenting the BRAF V600 mutation may not yield exchangeable information for statistical characterization of the effectiveness of the targeted therapy. Therefore, the MEM framework is implemented to measure the extent of basketwise heterogeneity and evaluate the effectiveness of the targeted therapy on the basis of its resultant multi-resolution smoothed posterior distributions. \cite{hobbs2018vemu} present a permutation study which extends the evaluation of heterogeneity to evaluate summaries of patient attributes reported in Table 1 of the aforementioned trial report. This case study reports posterior probabilities evaluating the evidence that the response probability for each organ-site exceeds the null rate of \code{p0} $=$ $0.25.$

The analysis can be reproduced by loading the \code{vemu_wide} data, which is included with the package. The data set includes the number of evaluable patients (column \code{evaluable}), the number of responding patients (column \code{responders}), and the associated baskets for the respective results (column \code{baskets}). The model is fit by passing these values to the \code{basket()} function along with an argument specifying the null response rate of 0.25 for evaluation of each basket. The results are shown by passing the fitted model object to the \code{summary()} function. Code to perform the analysis as well as produce the output is shown below.

\begin{Schunk}
\begin{Sinput}
data(vemu_wide)
vm <- basket(vemu_wide$responders, vemu_wide$evaluable,
             vemu_wide$baskets, p0 = 0.25)
summary(vm)
\end{Sinput}
\begin{Soutput}

-- The MEM Model Call ----------------------------------------------------------

mem_mcmc(responses = vemu_wide$responders, size = vemu_wide$evaluable, 
    name = vemu_wide$baskets, p0 = 0.25, mcmc_iter = 2e+05)

-- The Basket Summary ----------------------------------------------------------

The Null Response Rates (alternative is greater):
               NSCLC CRC (vemu) CRC (vemu+cetu) Bile Duct ECD or LCH   ATC
Null           0.250      0.250            0.25     0.250      0.250 0.250
Posterior Prob 0.971      0.002            0.00     0.229      0.968 0.896

Posterior Mean and Median Response Rates:
       NSCLC CRC (vemu) CRC (vemu+cetu) Bile Duct ECD or LCH   ATC
Mean   0.395      0.054           0.052     0.148      0.394 0.360
Median 0.392      0.046           0.045     0.096      0.391 0.363

Highest Posterior Density Interval with Coverage Probability 0.95:
            NSCLC CRC (vemu) CRC (vemu+cetu) Bile Duct ECD or LCH   ATC
Lower Bound 0.243      0.000           0.001     0.004      0.240 0.178
Upper Bound 0.553      0.129           0.121     0.402      0.553 0.555

Posterior Effective Sample Size:
  NSCLC CRC (vemu) CRC (vemu+cetu) Bile Duct ECD or LCH    ATC
 36.739     49.585          55.031    10.553     36.014 23.542

-- The Cluster Summary ---------------------------------------------------------

Cluster 1                                           
 "CRC (vemu)" "CRC (vemu+cetu)" "Bile Duct"
Cluster 2                           
 "NSCLC" "ECD or LCH" "ATC"

The Null Response Rates (alternative is greater):
               Cluster 1 Cluster 2
Null               0.250     0.250
Posterior Prob     0.077     0.945

Posterior Mean and Median Response Rates:
       Cluster 1 Cluster 2
Mean       0.085     0.383
Median     0.056     0.383

Highest Posterior Density Interval with Coverage Probability 0.95:
            Cluster 1 Cluster 2
Lower Bound     0.000     0.221
Upper Bound     0.316     0.556

Posterior Effective Sample Size:
 Cluster 1 Cluster 2
     9.528    30.779

\end{Soutput}
\end{Schunk}

\begin{figure}[htbp!]
\centering
  \begin{subfigure}[b]{0.45\textwidth}
    \centering
    \includegraphics[width=\textwidth]{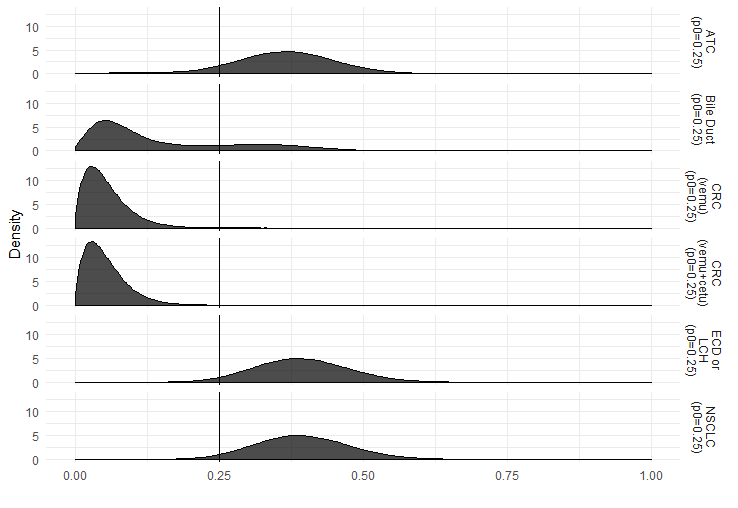}
    \caption{Posterior basket response density.}
    \label{cluster_post_density}
  \end{subfigure}
  \begin{subfigure}[b]{0.45\textwidth}
    \centering
    \includegraphics[width=\textwidth]{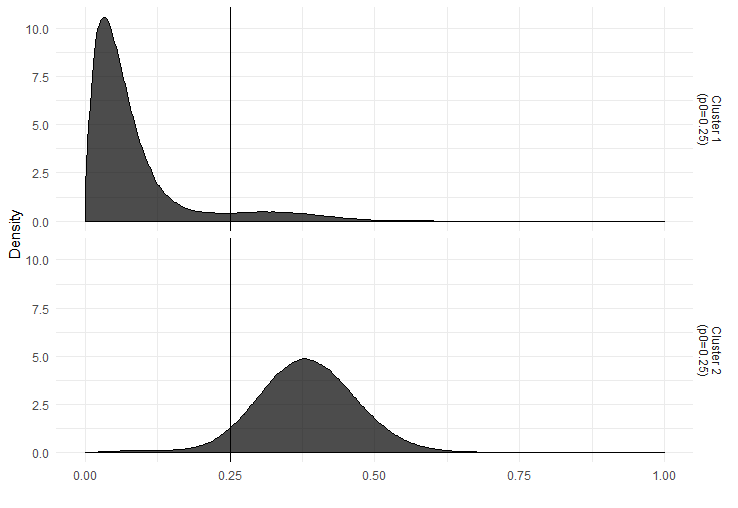}
    \caption{Posterior cluster response density.}
    \label{post_density}
  \end{subfigure}
  \caption{Posterior distributions of the MEM analysis.}
\end{figure}

Bayesian MEM analysis using the MCMC sampler with reference prior distribution for exchangeability identifies the most likely MEM to be comprised of two closed subgraphs (or meta-baskets). Cluster 1 consists of CRC.v with CRC.vc and BD, while cluster 2 is comprised of NSCLC, ED.LH, and ATC. Cluster 1 results in an estimated posterior mean response rate of $0.087.$ The posterior probability that baskets assigned to cluster 1 exceed the null response rate of $0.25$ is only $0.082.$ Conversely, attaining a posterior probability of $0.944$ and posterior mean of $0.382,$ indications identified in cluster 2 demonstrate more promising indications of activity. Figures \ref{cluster_post_density} and \ref{post_density} depict full posterior distributions of response probabilities for each basket and cluster produced by the \code{plot_density()} function.  

\begin{Schunk}
\begin{Sinput}
plot_density(vm, type = "basket")
plot_density(vm, type = "cluster")
\end{Sinput}
\end{Schunk}

\begin{figure}[htbp!]
\centering
\includegraphics{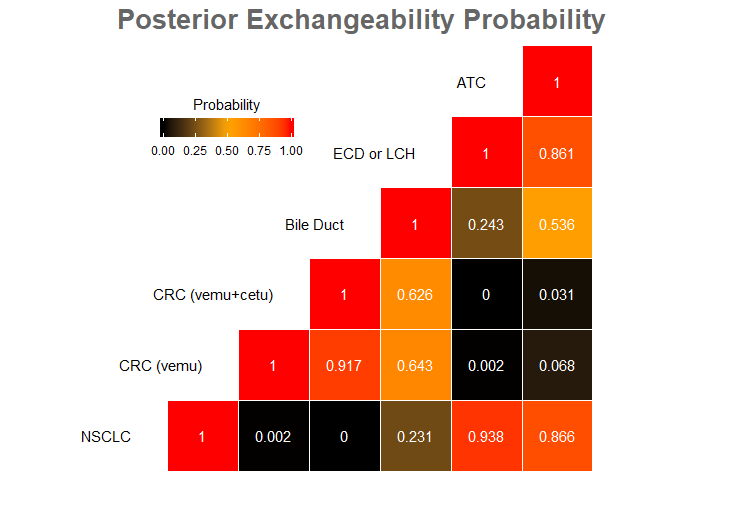}
\caption{The exchangeogram for the Vemurafenib study.}
\label{excgrm}
\end{figure}

The resultant posterior probability of each pairwise exchangeability relationship (PEP) is summarized with the \code{basket_pep()} function and depicted in Figure \ref{excgrm} by application of the \code{plot_pep()} function. The results demonstrate that the posterior exchangeability between the high-response baskets is higher than that of the lower responding baskets. For example, the posterior probability that vemurafenib offers identical effectiveness for NSCLC and ED.LH patients is 0.938. Similarly, the analysis resulted in PEPs of 0.86 for the pairwise relationships between NSCLC with ATC and ED.LH with ATC. The study provided strong support to conclude that vemurafenib is identically ineffective among CRC.v and CRC.vc subtypes with PEP $=$ $0.92.$ The effectiveness of BD was identified as marginally exchangeable with CRC.v and CRC.vc with PEP $=$ $0.64$ and $0.63,$ respectively. Conversely, both NSCLC and ED.LH resulted in PEPs of 0 for each CRC basket, demonstrating strong evidence of differential activity among these indications. Thus, definitive trials devised to estimate population-averaged effects should not expect these subtypes to comprise statistically exchangeable patients. 

%The result is of particular interest for the ATC basket, whose single-arm posterior probability of being greater than 0.25 is 0.614 under the Jeffrey's prior. In the MEM model, the ATC basket can borrow power from the similar NSCLC and ECD or LCH baskets, potentially providing more certainty the response rate is above the threshold. The analysis and it's summary is produced with the following. In the ``Basket Summary'' section in the ``Null Response Rates'' block, it is shown that the certainty that the ATC basket is above the 25\% threshold is now 93.3\% because the power that was borrowed in the analysis. Where the single basket trial had a sample size of 7, the MEM analysis has an effective sample size of almost 32. The posterior basket densities can visualized with the following code, whose output is shown in Figure \ref{post_density} and \ref{cluster_post_density}.
%
%The summary and visualizations also confirm our hypothesis there are two clusters based on the response similarity. One with relatively low response rates and one with a response rate above 25\%. %We can examine the posterior exchangeability between any two baskets with the following code, which retrieves the PEP matrix and produces the exchangeogram shown in Figure \ref{excgrm}.

\begin{Schunk}
\begin{Sinput}
basket_pep(vm)
plot_pep(vm$basket)
\end{Sinput}
\begin{Soutput}
                NSCLC CRC (vemu) CRC (vemu+cetu) Bile Duct ECD or LCH   ATC
NSCLC           1.000      0.002           0.000     0.231      0.938 0.866
CRC (vemu)      0.002      1.000           0.917     0.643      0.002 0.068
CRC (vemu+cetu) 0.000      0.917           1.000     0.626      0.000 0.031
Bile Duct       0.231      0.643           0.626     1.000      0.243 0.536
ECD or LCH      0.938      0.002           0.000     0.243      1.000 0.861
ATC             0.866      0.068           0.031     0.536      0.861 1.000
\end{Soutput}
\end{Schunk}

\section{Summary}

%Conventional strategies for trial design assume that all patients/indications enrolling into a particular study represent an ``average patient.'' 
With the emergence of molecularly targeted therapies, contemporary trials are devised to enroll potentially heterogeneous patient populations defined by a common treatment target. Consequently, characterization of subpopulation heterogeneity has become central to the design and analysis of clinical trials, in oncology in particular. By partitioning the study population into subpopulations that comprise potentially non-exchangeable patient cohorts, the basket design framework can be used to study treatment heterogeneity in a prospective manner. When applied in this context, the Bayesian multisource exchangeability model (MEM) methodology refines the estimation of treatment effectiveness to specific subpopulations. Additionally, the MEM inferential strategy objectively identifies which patient subpopulations should be considered exchangeable and to what extent.

This article introduced the R package \pkg{basket} as well as demonstrated its implementation for basket trial analysis using the MEM methodology. An oncology case study using data acquired from a basket trial was presented and used to demonstrate the main functionality of the package. \pkg{Basket} is the first available software package implementing Bayesian analysis with the MEM. The package is being actively maintained and used in ongoing trials.

\section*{}
\emph{Acknowledgements:} This work was partially supported by Amgen, Inc. as well as The Yale Comprehensive Cancer Center (P30CA016359), and The Case Comprehensive Cancer Center (P30 CA043703). 

\bibliography{references} 

\begin{thebibliography}{28}
\newcommand{\enquote}[1]{``#1''}
\providecommand{\natexlab}[1]{#1}
\providecommand{\url}[1]{\texttt{#1}}
\providecommand{\urlprefix}{URL }
\expandafter\ifx\csname urlstyle\endcsname\relax
  \providecommand{\doi}[1]{doi:\discretionary{}{}{}#1}\else
  \providecommand{\doi}{doi:\discretionary{}{}{}\begingroup
  \urlstyle{rm}\Url}\fi
\providecommand{\eprint}[2][]{\url{#2}}

\bibitem[{Berry \emph{et~al.}(2013)Berry, Broglio, Groshen, and
  Berry}]{berry2013ct}
Berry SM, Broglio KR, Groshen S, Berry DA (2013).
\newblock \enquote{Bayesian hierarchical modeling of patient subpopulations:
  Efficient designs of Phase {II} oncology clinical trials.}
\newblock \emph{Clinical Trials}, \textbf{10}(5), 720--734.

\bibitem[{Berry \emph{et~al.}(2010)Berry, Carlin, Lee, and
  M\"{u}ller}]{berry:2010}
Berry SM, Carlin BP, Lee JJ, M\"{u}ller P (2010).
\newblock \emph{Bayesian Adaptive Methods for Clinical Trials}.
\newblock Chapman and Hall/CRC Press, Boca Raton, FL.

\bibitem[{Campbell and Yau(2017)}]{mfa}
Campbell KR, Yau C (2017).
\newblock \enquote{Probabilistic modeling of bifurcations in single-cell gene
  expression data using a Bayesian mixture of factor analyzers.}
\newblock \emph{Wellcome open research}, \textbf{2}.
\newblock \doi{10.12688/wellcomeopenres.11087.1}.
\newblock \urlprefix\url{http://dx.doi.org/10.12688/wellcomeopenres.11087.1}.

\bibitem[{Chen \emph{et~al.}(2019)Chen, Hobbs, Kaizer, and Kane}]{basket}
Chen N, Hobbs B, Kaizer A, Kane MJ (2019).
\newblock \emph{basket: Basket Trial Analysis}.
\newblock R package version 0.9.2.

\bibitem[{Cunanan \emph{et~al.}(2017{\natexlab{a}})Cunanan, Gonen, Shen, Hyman,
  Riely, Begg, and Iasonos}]{doi:10.1200/JCO.2016.69.9751}
Cunanan KM, Gonen M, Shen R, Hyman DM, Riely GJ, Begg CB, Iasonos A
  (2017{\natexlab{a}}).
\newblock \enquote{Basket Trials in Oncology: A Trade-Off Between Complexity
  and Efficiency.}
\newblock \emph{Journal of Clinical Oncology}, \textbf{35}(3), 271--273.
\newblock \doi{10.1200/JCO.2016.69.9751}.
\newblock PMID: 27893325, \eprint{https://doi.org/10.1200/JCO.2016.69.9751},
  \urlprefix\url{https://doi.org/10.1200/JCO.2016.69.9751}.

\bibitem[{Cunanan \emph{et~al.}(2017{\natexlab{b}})Cunanan, Iasonos, Shen,
  Hyman, Riely, G{\"o}nen, and Begg}]{Cunananetal17specifying}
Cunanan KM, Iasonos A, Shen R, Hyman DM, Riely GJ, G{\"o}nen M, Begg CB
  (2017{\natexlab{b}}).
\newblock \enquote{Specifying the True-and False-Positive Rates in Basket
  Trials.}
\newblock \emph{JCO Precision Oncology}, \textbf{1}, 1--5.

\bibitem[{Durie \emph{et~al.}(2006)Durie, Harousseau, Miguel, Blade, Barlogie,
  Anderson, Gertz, Dimopoulos, Westin, Sonneveld \emph{et~al.}}]{durie2006}
Durie BG, Harousseau J, Miguel J, Blade J, Barlogie B, Anderson K, Gertz M,
  Dimopoulos M, Westin J, Sonneveld P, \emph{et~al.} (2006).
\newblock \enquote{International uniform response criteria for multiple
  myeloma.}
\newblock \emph{Leukemia}, \textbf{20}(9), 1467.

\bibitem[{Eisenhauer \emph{et~al.}(2009)Eisenhauer, Therasse, Bogaerts,
  Schwartz, Sargent, Ford, Dancey, Arbuck, Gwyther, Mooney
  \emph{et~al.}}]{eisenhauer2009}
Eisenhauer EA, Therasse P, Bogaerts J, Schwartz LH, Sargent D, Ford R, Dancey
  J, Arbuck S, Gwyther S, Mooney M, \emph{et~al.} (2009).
\newblock \enquote{New response evaluation criteria in solid tumours: revised
  RECIST guideline (version 1.1).}
\newblock \emph{European journal of cancer}, \textbf{45}(2), 228--247.

\bibitem[{Freidlin and Korn(2013)}]{freidlin2013ccr}
Freidlin B, Korn E (2013).
\newblock \enquote{Borrowing Information across Subgroups in Phase {II} Trials:
  Is It Useful?}
\newblock \emph{Clinical Cancer Research}, \textbf{19}, 1326--1334.

\bibitem[{Gelman \emph{et~al.}(2013)Gelman, Carlin, Stern, and
  Rubin}]{gelman:2013}
Gelman A, Carlin JB, Stern HS, Rubin DB (2013).
\newblock \emph{Bayesian Data Analysis}.
\newblock 3rd edition. Chapman and Hall/CRC Press, Boca Raton, FL.

\bibitem[{Hobbs \emph{et~al.}(2018)Hobbs, Kane, Hong, and
  Landin}]{hobbs2018vemu}
Hobbs B, Kane M, Hong D, Landin R (2018).
\newblock \enquote{Statistical challenges posed by uncontrolled master
  protocols: sensitivity analysis of the vemurafenib study.}
\newblock \emph{Annals of Oncology}, \textbf{29}(12), 2296--2301.

\bibitem[{Hobbs \emph{et~al.}(2013)Hobbs, Carlin, and Sargent}]{hobbs2013}
Hobbs BP, Carlin BP, Sargent DJ (2013).
\newblock \enquote{Adaptive adjustment of the randomization ratio using
  historical control data.}
\newblock \emph{Clinical Trials}, \textbf{10}, 430--440.

\bibitem[{Hobbs and Landin(2018)}]{hobbs2018monitor}
Hobbs BP, Landin R (2018).
\newblock \enquote{Bayesian basket trial design with exchangeability
  monitoring.}
\newblock \emph{Statistics in medicine}, \textbf{37}(25), 3557--3572.

\bibitem[{Hyman \emph{et~al.}(2015)Hyman, Puzanov, Subbiah, Faris, Chau, Blay,
  Wolf, Raje, Diamond, Hollebecque \emph{et~al.}}]{hyman2015}
Hyman DM, Puzanov I, Subbiah V, Faris JE, Chau I, Blay JY, Wolf J, Raje NS,
  Diamond EL, Hollebecque A, \emph{et~al.} (2015).
\newblock \enquote{Vemurafenib in multiple nonmelanoma cancers with BRAF V600
  mutations.}
\newblock \emph{New England Journal of Medicine}, \textbf{373}(8), 726--736.

\bibitem[{Kaizer \emph{et~al.}(2018)Kaizer, Hobbs, and
  Koopmeiners}]{kaizer2018}
Kaizer AM, Hobbs BP, Koopmeiners JS (2018).
\newblock \enquote{A multi-source adaptive platform design for testing
  sequential combinatorial therapeutic strategies.}
\newblock \emph{Biometrics}, \textbf{74}(3), 1082--1094.

\bibitem[{Kaizer \emph{et~al.}(2017)Kaizer, Koopmeiners, and
  Hobbs}]{kaizer2017}
Kaizer AM, Koopmeiners JS, Hobbs BP (2017).
\newblock \enquote{Bayesian hierarchical modeling based on multisource
  exchangeability.}
\newblock \emph{Biostatistics}, \textbf{19}(2), 169--184.

\bibitem[{Mo(2018)}]{iSeq}
Mo Q (2018).
\newblock \emph{iSeq: Bayesian Hierarchical Modeling of ChIP-seq Data Through
  Hidden Ising Models}.
\newblock R package version 1.34.0.

\bibitem[{Murray \emph{et~al.}(2015)Murray, Hobbs, and Carlin}]{murray2015}
Murray TA, Hobbs BP, Carlin BP (2015).
\newblock \enquote{Combining nonexchangeable functional or survival data
  sources in oncology using generalized mixture commensurate priors.}
\newblock \emph{Annals of Applied Statistics}, \textbf{9}(3), 1549--1570.

\bibitem[{Nia and Davison(2012)}]{bclust}
Nia VP, Davison AC (2012).
\newblock \enquote{High-Dimensional Bayesian Clustering with Variable
  Selection: The {R} Package {bclust}.}
\newblock \emph{Journal of Statistical Software}, \textbf{47}(5), 1--22.
\newblock \urlprefix\url{http://www.jstatsoft.org/v47/i05/}.

\bibitem[{{R Core Team}(2019)}]{Rcore}
{R Core Team} (2019).
\newblock \emph{R: A Language and Environment for Statistical Computing}.
\newblock R Foundation for Statistical Computing, Vienna, Austria.
\newblock \urlprefix\url{https://www.R-project.org/}.

\bibitem[{Samb \emph{et~al.}(2015)Samb, Khadraoui, Belleau, Desch\^enes,
  Lakhal-Chaieb, and Droit}]{RJMCMCNucleosomes}
Samb R, Khadraoui K, Belleau P, Desch\^enes A, Lakhal-Chaieb L, Droit A (2015).
\newblock \enquote{Using informative Multinomial-Dirichlet prior in a t-mixture
  with reversible jump estimation of nucleosome positions for genome-wide
  profiling.}
\newblock \emph{Statistical Applications in Genetics and Molecular Biology},
  \textbf{14}.
\newblock \doi{10.1515/sagmb-2014-0098}.
\newblock
  \urlprefix\url{http://www.degruyter.com/view/j/sagmb.ahead-of-print/sagmb-2014-0098/sagmb-2014-0098.xml}.

\bibitem[{Savage \emph{et~al.}(2018)Savage, Cooke, Darkins, and Xu}]{BHC}
Savage R, Cooke E, Darkins R, Xu Y (2018).
\newblock \emph{BHC: Bayesian Hierarchical Clustering}.
\newblock R package version 1.34.0.

\bibitem[{Scharpf \emph{et~al.}(2009)Scharpf, Tjelmeland, Parmigiani, and
  Nobel}]{XDE}
Scharpf RB, Tjelmeland H, Parmigiani G, Nobel A (2009).
\newblock \enquote{A Bayesian model for cross-study differential gene
  expression.}
\newblock \emph{JASA}.
\newblock \urlprefix\url{10.1198/jasa.2009.ap07611}.

\bibitem[{Sharifi-Malvajerdi \emph{et~al.}(2019)Sharifi-Malvajerdi, Zhu,
  Fogarty, Fay, Fairhurst, Flegg, Stepniewska, and Small}]{bhrcr}
Sharifi-Malvajerdi S, Zhu F, Fogarty CB, Fay MP, Fairhurst RM, Flegg JA,
  Stepniewska K, Small DS (2019).
\newblock \enquote{Malaria parasite clearance rate regression: an R software
  package for a Bayesian hierarchical regression model.}
\newblock \emph{Malaria Journal}, \textbf{18}(1), 4.
\newblock ISSN 1475-2875.
\newblock \doi{10.1186/s12936-018-2631-8}.
\newblock \urlprefix\url{https://doi.org/10.1186/s12936-018-2631-8}.

\bibitem[{Stocco(2014)}]{stocco2014}
Stocco A (2014).
\newblock \enquote{Coordinate-Based Meta-Analysis of fMRI Studies with R.}
\newblock \emph{R Journal}, \textbf{6}(2).

\bibitem[{Thall \emph{et~al.}(2003)Thall, Wathen, Bekele, Champlin, Baker, and
  Benjamin}]{thall2003sim}
Thall P, Wathen J, Bekele B, Champlin R, Baker L, Benjamin R (2003).
\newblock \enquote{Hierarchical Bayesian approaches to phase {II} trials in
  diseases with multiple subtypes.}
\newblock \emph{Statistics in Medicine}, \textbf{22}, 763--780.

\bibitem[{{Yang Xiang} \emph{et~al.}(2013){Yang Xiang}, Gubian, Suomela, and
  Hoeng}]{gensa}
{Yang Xiang}, Gubian S, Suomela B, Hoeng J (2013).
\newblock \enquote{Generalized Simulated Annealing for Efficient Global
  Optimization: the {GenSA} Package for {R}.}
\newblock \emph{The R Journal Volume 5/1, June 2013}.
\newblock
  \urlprefix\url{https://journal.r-project.org/archive/2013/RJ-2013-002/index.html}.

\bibitem[{Zhang and Zhang(2018)}]{clinicaltrialstask}
Zhang E, Zhang HG (2018).
\newblock \enquote{{CRAN Task View: Clinical Trial Design, Monitoring, and
  Analysis}.}
\newblock \url{https://CRAN.R-project.org/view=ClinicalTrials }.
\newblock Version~2018-06-18.

\end{thebibliography}

\end{document}